# COMPARATIVE STUDY AND ANALYSIS OF VARIABILITY TOOLS

Mahendra Reddy Bhumula
School of Computing, IT and Engineering
University of East London
London, UK
prince4umahi@gmail.com

*Abstract-* The dissertation provides a comparative analysis of a number of variability tools currently in use. It serves as a catalogue for practitioners interested in the topic. We compare a range of modelling, configuring, and management tools for product line engineering. The tools surveyed are compared against the following criteria: functional, non-functional, governance issues and Technical aspects. The outcome of the analysis is provided in tabular format.

## 1.0 Introduction

Variability management in software product lines encompasses the activities of explicitly representing variability in software artefacts throughout the lifecycle, managing the dependencies among different variability's and supporting the instantiations of those variability's. This will involve in a complex and challenging task to be supported by appropriate approaches, techniques, and tools.

We have given a vital role for variability management in Software Product Line Engineering. There is great deal of research included in this field. So many approaches and tools have been developed with the basic aim of supporting mostly all the tasked that involved in variability Management at different stages of product line's life.

In this survey we used tools which supporting all phases of life cycle like analysis, design, and implementation. With this also provided the approach of the specific tool in the approach like configuring, modelling, and management. Including GEARS (1) software from BigLever inc, COVAMOF (2), FAMA Tool Suite (9) etc have up to 14 tools included in this total survey.

This report gives the information about the tools which are supporting and not supporting to functional, non-functional, and governance issues. By using the information from this report people from industry can pick their tool with ease without any doubt. It is a fact that without the knowledge of tool properties it is very difficult to pick a tool for use. This report gives a comparative analysis of tools which includes the tools of low-end to high-end. The comparison is based on different properties.

### 1.1 Scope of the Work:

The Project work aims to develop a brief conceptual understanding of variability management. The main focus is to identify the variability tools and to perform the comparative analysis of these tools based on the functional, non-functional, governance and technical aspects. The work is maximised to an extent of comparing more than Ten Tools. The final work is to propose the best tools for industries or individuals to pick one effective tools that meets all the criteria defined in comparison.

**1.2 Achievable Goals:**

- To Understand & Analyse the conceptual work of variability management in software product line engineering.
- To perform an extensive search of variability management tools for the purpose of analysis.
- To analyse the tools based on different aspects mentioned in scope of the work.
- To report the results of analysis in well understood document format.

**1.3 Required Resources:**

In order to perform the study and analysis of the proposed work an access to the material related to the subject published in various books, journals, and online news and technical papers is necessary. In order to collect the material the required set of resources are:

A personal computer with minimum configuration (Can be used to browse the internet),

An access to library for books,

An access to the digital library to search for published journals.

## 2.0 Survey of the Variability Management Tools

Managing the variability became a necessary business requirement in software product line. It is due to the fact that, the current trends in variability of moving hardware to software lead the industries to postpone the decisions of designing aspects till it become economically feasible. As mentioned in the introductory part, this project focus mainly on identifying different types of variability tools and to perform the analysis of these tools on the basis of functional, non-functional, governance and technical aspects. Some of the tools that are identified during the online tool survey are mentioned in this chapter with brief explanation of each. Totally 14 (Numeric) Tools are identified and are mentioned in the list below:

**Tools for survey**

1. GEARS Tool
2. COVAMOF
3. VMWT
4. AHEAD
5. CONSUL
6. Feature Modelling Tool
7. Pure::variants
8. Feature Plug-in for eclipse
9. FAMA Tool Suite
10. KUMBANG
11. XToF
12. PLUSEE
13. DecisionKing
14. BVR Tool

## 3.0 GEARS Tool:

Gears (1) is a commercial SPL(software product line) development tool developed by **BigLever** Inc (15) and enables the modelling of optional and varying features which is used to differentiate the products in portfolio.

The Gears feature model uses high level typing includes (sets, enumeration, records, Boolean, integer, float, character, string) showing difference between "features" at the domain modelling level and "variation points" at the implementation level (source code, requirements, test cases, documentation).

In Gears,

- **Set** types allow the selection of optional objects.
- **Enumeration** types allow selection of one and only alternative.
- **Boolean** represent singular options.
- **Record** represents mandatory lists of features.

Gears variation points are inserted to support implementation level variation. Components with Gears variation points become reusable core assets that are automatically composed and configured into product instances. The workers in Gears given us a conventional way on Gears assets, with the expectation of implementing the variation points to support the feature model variations that are in the scope of their first asset.

If we see the dependency section in Gears are expressed as relational assertions. They used very simple conventional require and excludes dependencies. Variation points and feature models are fully user programmable to arbitrary levels of sophistication and complexity. The Gears approach defines product feature profiles for each product and selects the desired choices in the feature model. A product configurator automatically produces the individual products in the portfolio by assembling the assets and customizing the variation points within those assets to produce a particular product according to the feature profile. Gears modules can be mapped to any existing modularity capabilities in software. Basically Gears models can be composed into which can be treated as standalone "product lines". These product lines can be composed from modules and other nested product lines. Aspect- oriented features are captured in Gears "mix-ins", which allow crosscutting features to be imported into one or more modules for use in implementation variation points in those modules. The tool supports also the definition of hierarchical product lines by nesting one product line into another (1).

There are two types of views and editor styles are supported and can be switched dynamically

    a) syntactically and semantically well-defined text view
    b) context-sensitive structural tree view

Basically in Gears uses file and text based configuration and composition. This language-independent approach allows users to translation legacy variation as well as implements new variations. For all the above and multiple binding times in one product line will be supported by Gears. Indirectly through statically instantiated configuration files or database settings Gears typically influence the runtime behaviour for runtime binding with these able to set dynamically making feature selection at runtime. Due to quickly adaption of a software mass customization for product line, Gears enable organisations to use the software mass customization technology in a easy way. Proactive reactive and extractive approaches can be used depending of each particular organisation, but they are not mutually exclusive. Gears have been used in systems with millions of LoC with no prevented limitation on scalability (1).

**4.0 COVAMOF**:

COVAMOF (2,32) (conIPF variability modelling framework)

It is a variability modelling approach to represent variation points and variants on all abstractions layers, supports the modelling of relations between dependencies, provides traceability, and a hierarchical organization of variability (2).

There will be five different variation points can be supported in COVAMOF

There are

1. optional
2. alternative
3. optional-variant
4. variants
5. value

The first variation point refers to the selection (zero or more) from the one or more associated variants. The COVAMOF variability view (CVV) represents the view of the variability for the product family artefacts and unifies the variability on all layers of abstraction. The CVV models the dependencies that occur in industrial product families to restrict the building of one or more variation points.

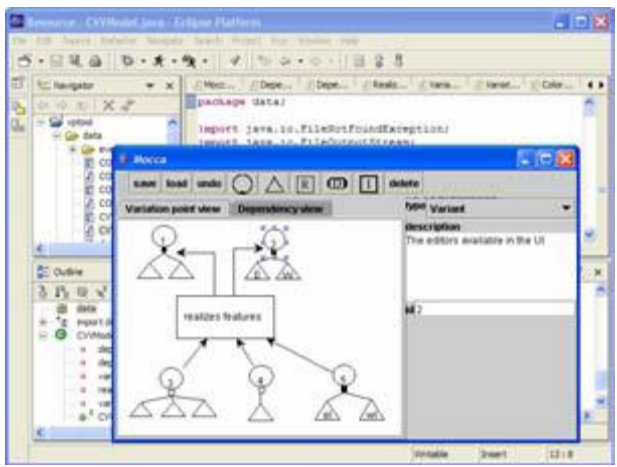

(2)

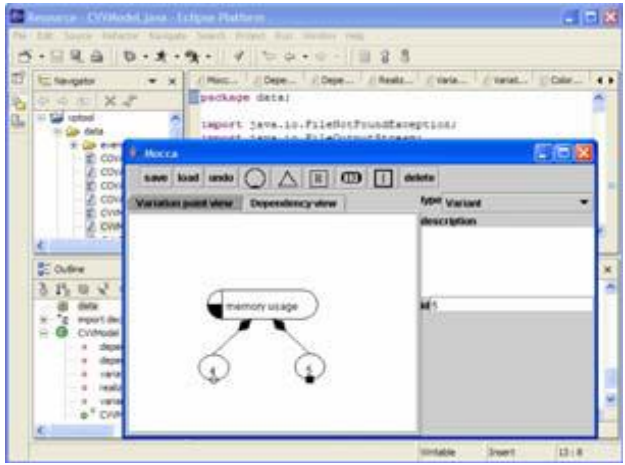

(2)

Simple dependencies are expressed by a Boolean expression, and CVV specifies a function valid to indicate whether a dependency is violated or not. In addition to the Boolean, dependencies and constraints can also contain integer values, with operators like the ADD, SUBSTRACT, etc. Boolean and numerical values are used together in operators like the GREATER THAN, where numerical values are the input and Boolean values are the output, complex dependencies are defined in COVAMOF as dynamically analyzable dependencies and CVV contains for each dynamically analyzable dependency the below stated properties(2).

- **Aspect**: Each dependency is associated with an aspect that can be expressed by a real value.
- **Valid range**: The dependency specifies a function to (true, false) indicating whether a value is acceptable.
- **Associations**: The CVV distinguishes three types of associations for dynamic dependencies, which are: predictable, directional and unknown.

For communication between tools COVAMOF provides graphical representation and XML representation. For multiple views of CVV and COVAMOF variability view Mocca tool has been developed to manage. Mocca supports the management of the CVV from the variation point view and the dependency view. Mocca is implemented in Java as extension to the eclipse 3.0 platform. Some recent improvements to COVAMOF-VS tool suite, which is a set of add-ns for Microsoft visual studio.NET. The COVAMOF-VS provides two main graphical views, that is variation point view and the dependency view, as a way to maintain an integrated variability model. Finally, specific plug-ins can be added for supporting different variability implementation mechanisms (2).

## 5.0 VMWT (Variability modelling web Tool):

VMWT is a research prototype developed at the university Juan Carlos of Madrid. This is first prototype (http://triana.escet.urjc.es/VMWT/) is a web-based tool built with PHP and Ajax and running over Apache 2.0. VMWT stores and manages variation points and variants following a product line approach and enables to create product line projects for which a set of reusable existing assets can be associated. Before configuring a particular code component and we can specify numeric values (quantitative values), ranges of values or a enumerated list can be specifies. Once all the variants have been added, the variation points will be added to the code components.

VMWT supports dependency rules and constraints for the variation points and variant already defined. The following Boolean relationships are allowed: AND, OR, XOR and NONE. In addition, more complex dependencies can be defined, such as requires and excludes. The tool allows constraint and dependency checking and we complete the number of allowed configurations. This is quite useful when it is needed to estimate the cost of the products to be engineered. Finally, a FODA tree is visualized for selecting the options for each product and the selected configuration is then displayed to the user. The variation points and variants selected are included in a file attached to each code component documentation of the product line can be automatically generated as PDF documents.

### 6.0 AHEAD Tool Suite (Algebraic Hierarchical Equations for Application Design)

The AHEAD (Algebraic Hierarchical Equations for application Development) Tool Suite (AHEAD TS) was developed to support the development of product lines using compositional programming techniques(34).

AHEAD TS has been used in distinct domains

  i. To produce applications where feature and variations are used in the production process(35).
  ii. To produce a product line of portlets.

The production process in software product line requires the use of features that have to be modelled as first-class entities. AHEAD distinguishes between "product features" and "built-in features". The former characterizes the product as such. The latter refers to variations on the associated process. The production processes are specified in using *Ant,* a popular scripting language from the java community. AHEAD uses a step-wise refinement process based on the GenVoca methodology for incrementally adding features to the products belonging to a system family.

The refinements supported by AHEAD (20) are packaged in layers. The base layer contains the base artefacts with specific features. The AHEAD production process shows differences between two stages. The intra-layer production process specifies the tasks for producing a set of artefacts within a layer or upper layers. The inter-layer production process defined how layers should be intertwined to obtain the final product. An extension to AHEAD is described in(36) and a Tool called XAK was developed for composing base and refinement artefacts in XML format. ATS was re-factored into features to allow the integration with XAK. The feature refactoring approach used in XAK decomposes legacy applications into set of feature modules which can be added to a product line. AHEAD doesn't require manual intervention during the derivation process.

### 7.0 CONSUL Based Tools:

Variability management tools have to be used by two different classes of users. The first class is formed by the developers of these variable artefacts. As a complete tool chain, CONSUL (5,33) supports both classes. The modular implementation of CONSUL allows flexible combining of the required services and user interfaces to build different tools.

The current application family consists of following three different tools

1. Consul@GUI
2. Consul@CLI
3. Consul@Web

**7.1 Consul@GUI     :**

The main application for developers is Consul@GUI is an interactive modelling tool for CONSUL models. It allows creating and editing the models but can also be used in the deployment of the developed software for generating the customized software. The screenshot represents the Consul@GUI of cosine domain with several features selected.

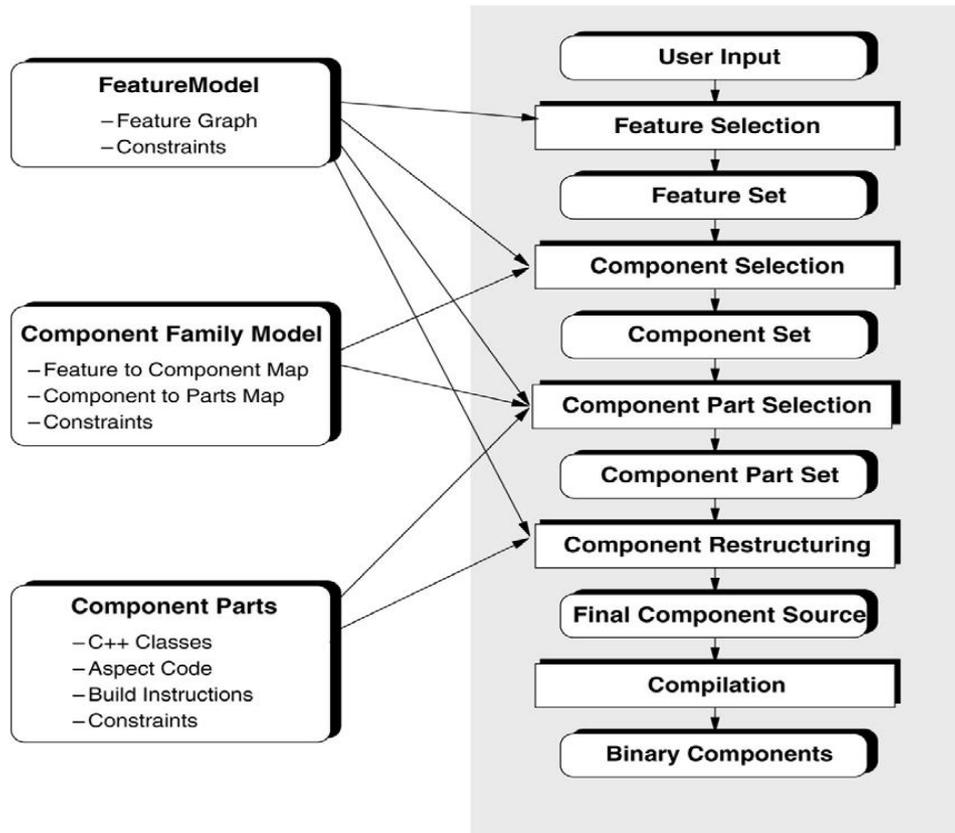

**Consul process Overview (5)**

The configuration is not valid, since there is still an open alternative. This is indicated by the background colours of the two features. Once a valid configuration has been found, the generation process can be started.

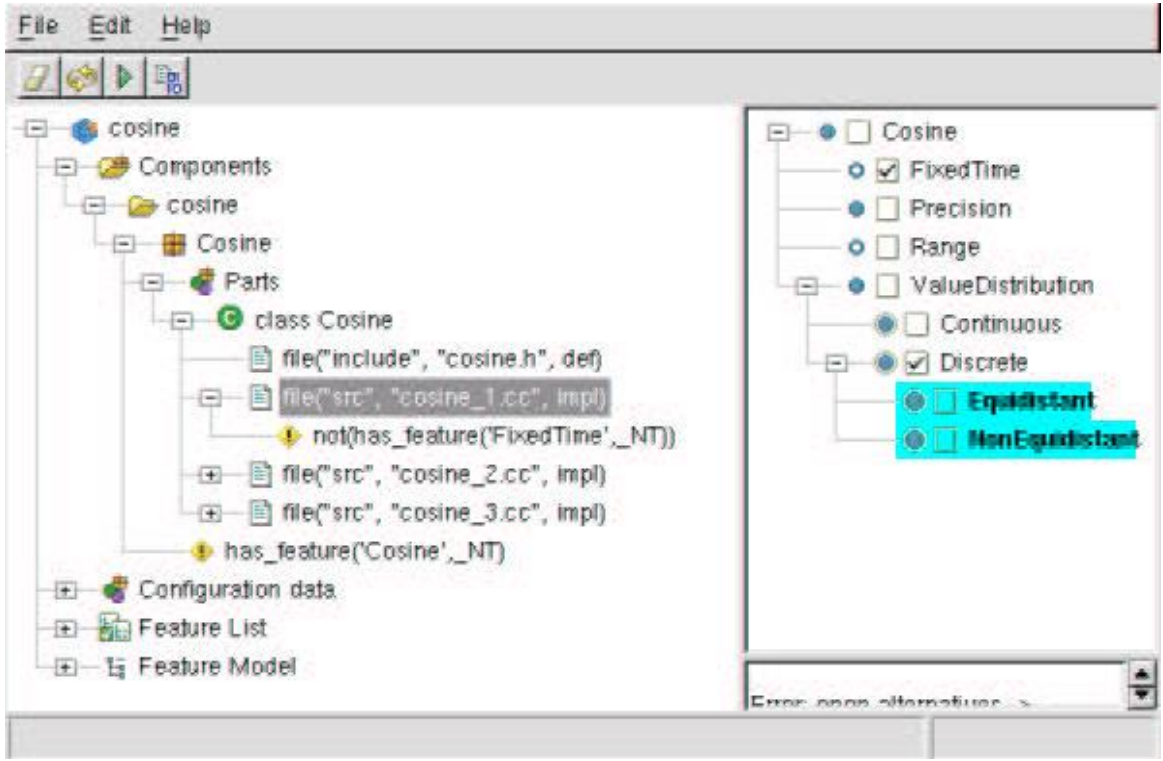

**Consul@GUI (5)**

### 7.2 Consul@CLI:
Based on CONSUL a customization tool with a command line interface has been built as well. This tool can be used e.g. together with make to provide automated customization when re-building a software system.

### 7.3 Consul@Web:
It is also possible to make software customization available via web browsers. A demonstration on a Java applet can be found in pure-systems website. It allows the configuration, building and downloading of pure via an Java-enabled web browser.

## 8.0 Feature Modelling tool:

This Feature modeling tool(21) will allow us to crate feature models from inside visual studio IDE. By using this tool we can visualize a) indented list b) Tree structure

Below figure is hierarchy of the feature modelling tool (6) where nodes in the left window represents the features. The central window represents the modeller's design where it is allowed to add/modify/delete features. It is a tree type representation where the links symbolize the hierarchy component and the nodes symbolize features(22).

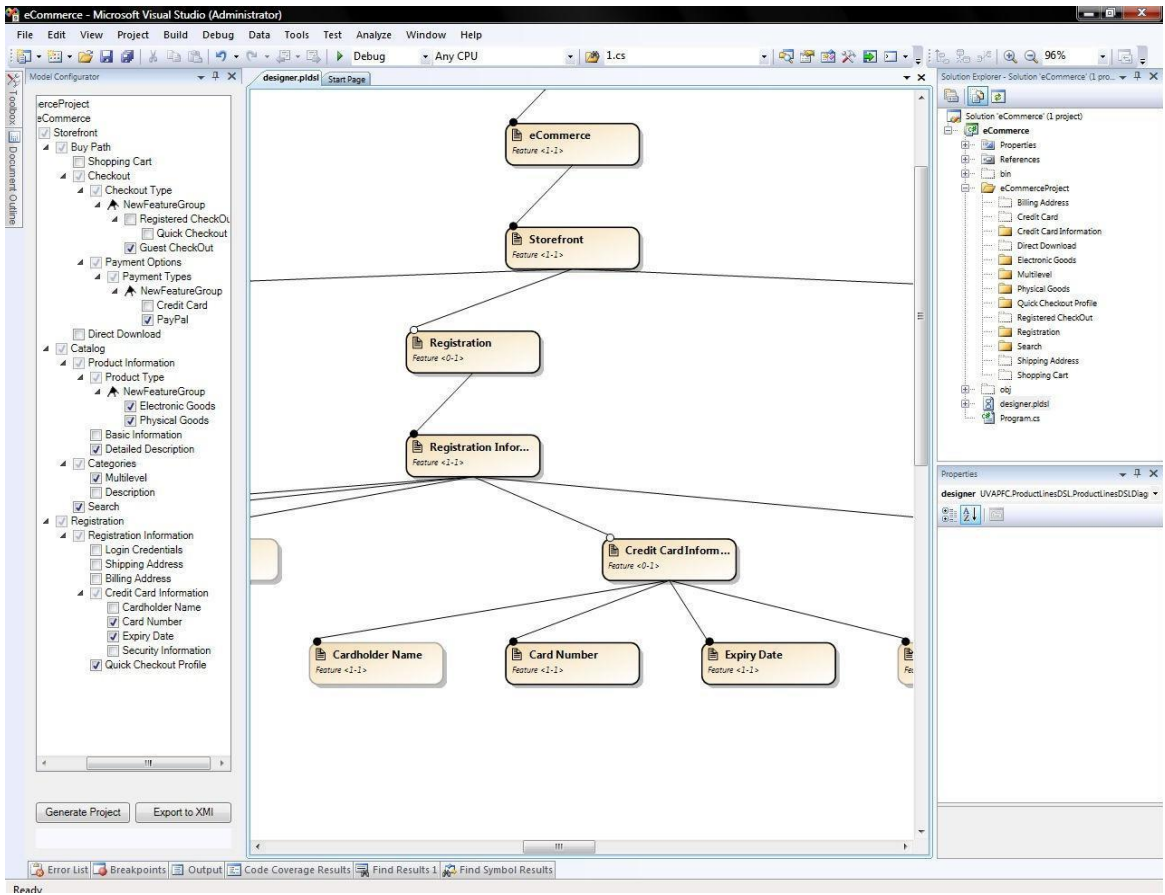

## 8.1 Feature model plug-in for eclipse

This is an eclipse plug-in (8,19) that represents feature models based on an indented list. It is very similar to the above tool, the nodes of the visuals represent the features also the or – groups and the alternative groups are placed under a new node that informs about the type of the grouping.

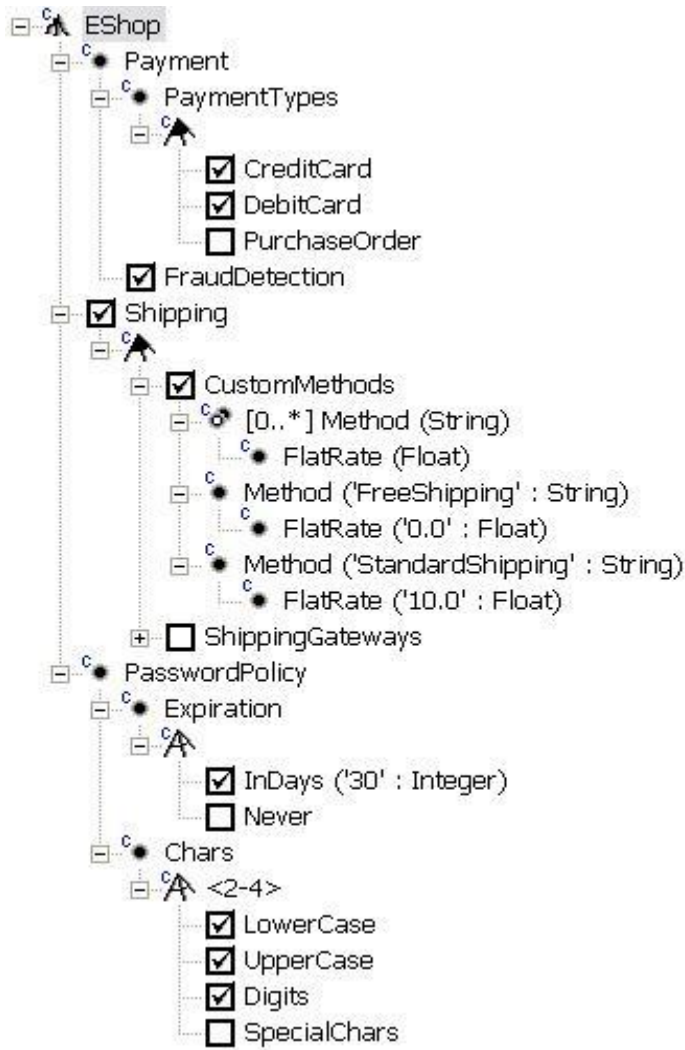

Configuration process fig-1

Constraints Evaluated from Configuration fig-2

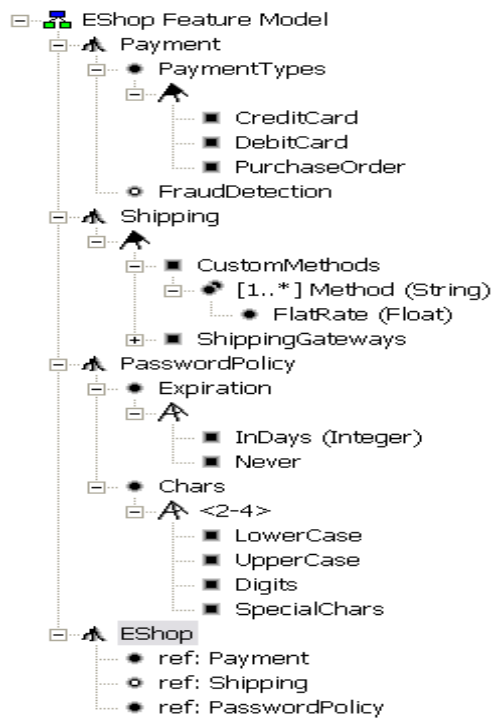

Example of Feature model in Editor view fig -3

This tool supports cardinality-based feature modelling, specialization of feature diagrams, and configuration based on feature diagrams. This is an eclipse plug-in for feature modelling. This an ability of bringing the eclipse (16) platform closer to software- product line and generative development communities, providing tool support for feature modelling as an Ecplise plug-in is particularly attractive with the below reasons(OOPSLA'04). Initially, integration feature modelling as a part of a development environment helps to optimally support modelling variability in different artefacts.

**Example of Feature model in editor view:** When the user clicks on the feature, an auxiliary window shows the information about the node. In addition to that feature dependencies are not available in this model.

## 9.0 Pure::variants:

Pure::Variants (7) is a commercial Tool supporting feature modelling and configuration using tree-view rendering. In other words, pure::Variants does not support cloning, pure::Variants allows modelling global constraints between features and it offers interactive, constraints-based configuration using a prolog-based constraint solver. It is also a feature modelling tool which is created by pure systems GmBh set up in 2001. Basically it is an eclipse application, an open source community whose projects are focused on building an open development platform comprised of extensible frame works and the main functionality of it is to be used a frame work for the design of product line architectures(23).

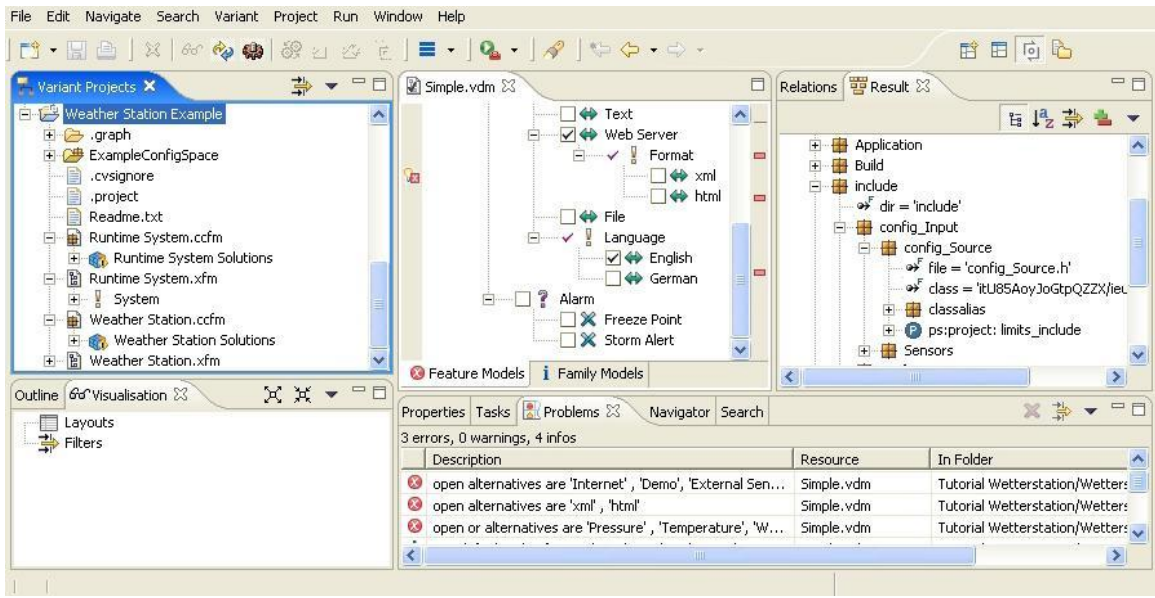

It is one of the representations of the previous tool, the items are situated as nodes in an indented list. Each check box placed near each component is used to configure a product line from the feature model. Thus the user is allowed to display a final result of a product line, if he or she selects some configuration by the use of these check-boxes (7).

Pure::variants adds the possibility to represent the model by graph visualization, although some common editing operations like editing, deletion are supported by the tool.

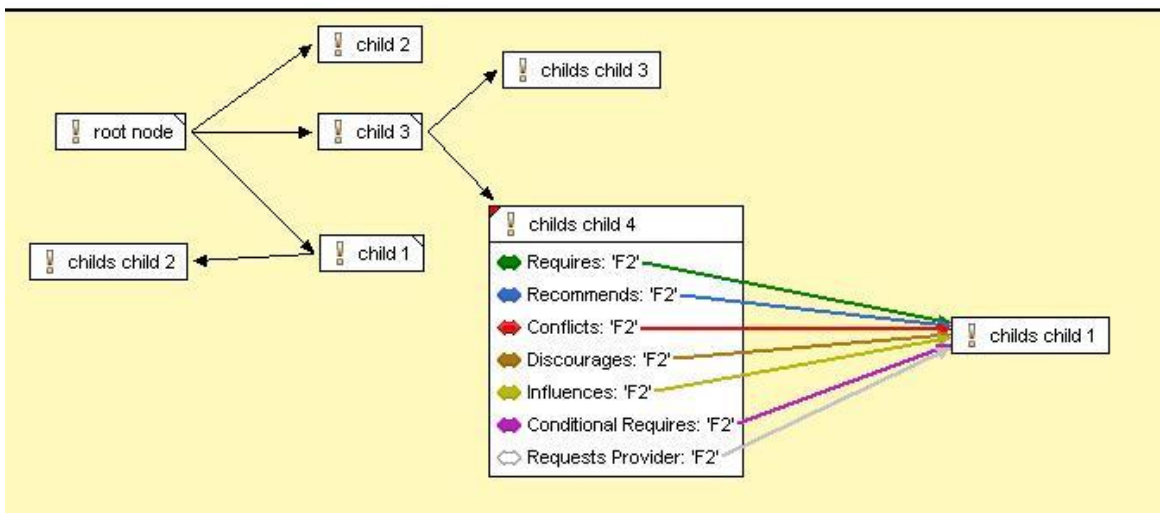

Pure::variants graph visualization figure (7).

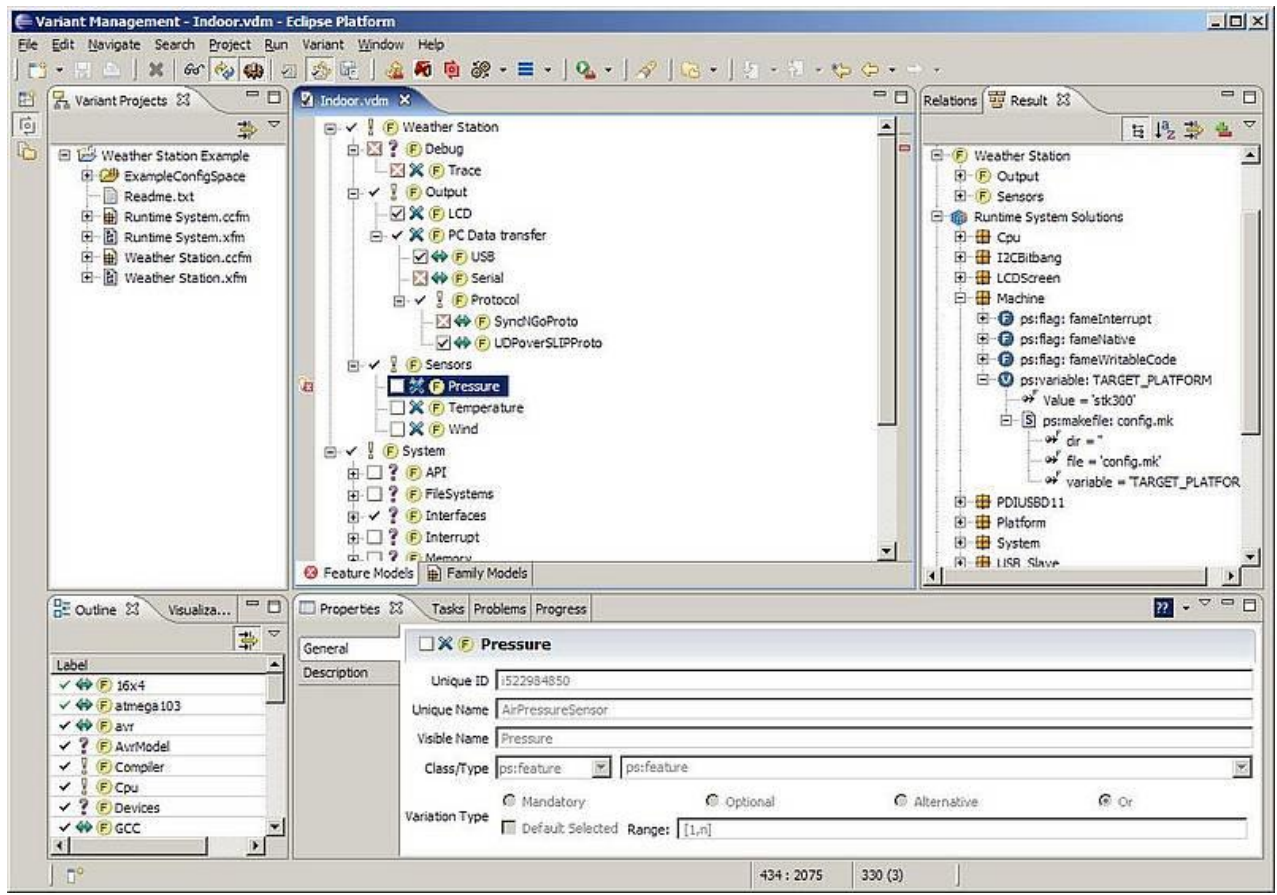

**Configuration editor**

## 10.0 FAMA tool suite:

This is a tool for the automated analysis of the variability models. The application provides an extensible framework for easily reading variability models, and automating the configuration of a final product. For the analysis and edition of Feature models FAMA (9) has been implemented as a complete tool. FAMA supports cardinality based feature modelling, export/import of Feature models from the XML and XMI and analysis operations of Feature models.

As the majority of the feature modelling applications, FAMA Tool suite uses GUI as a representation of the model. The difference in this case lies in the process of modelling; the user has to develop the structure of the feature model writing it in an XML. Then the tools read the document and visualize the content of it like in the figure shown below allowing the user to interact with the representation.

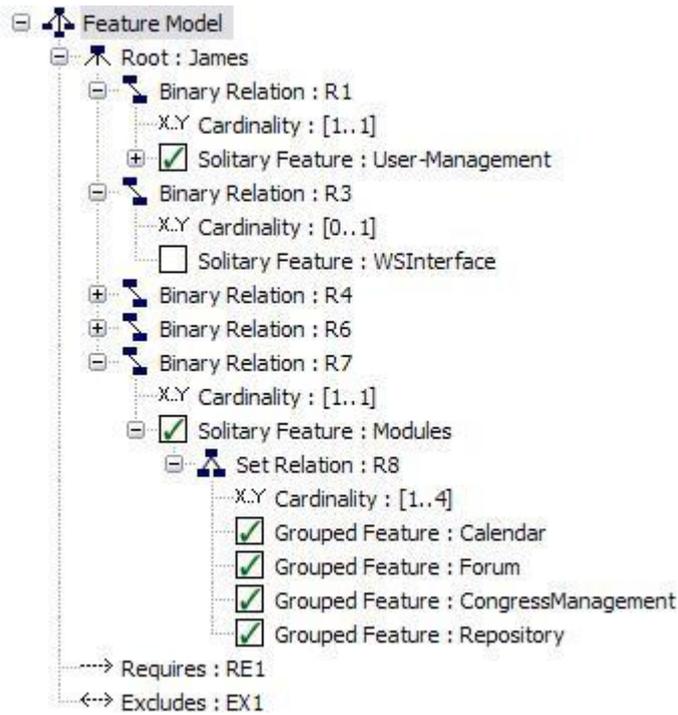
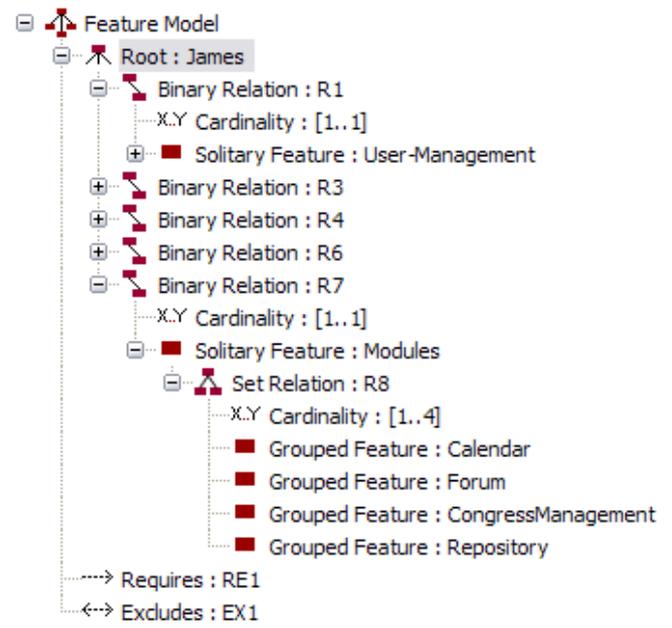

**Analysis view**     **Modelling view**

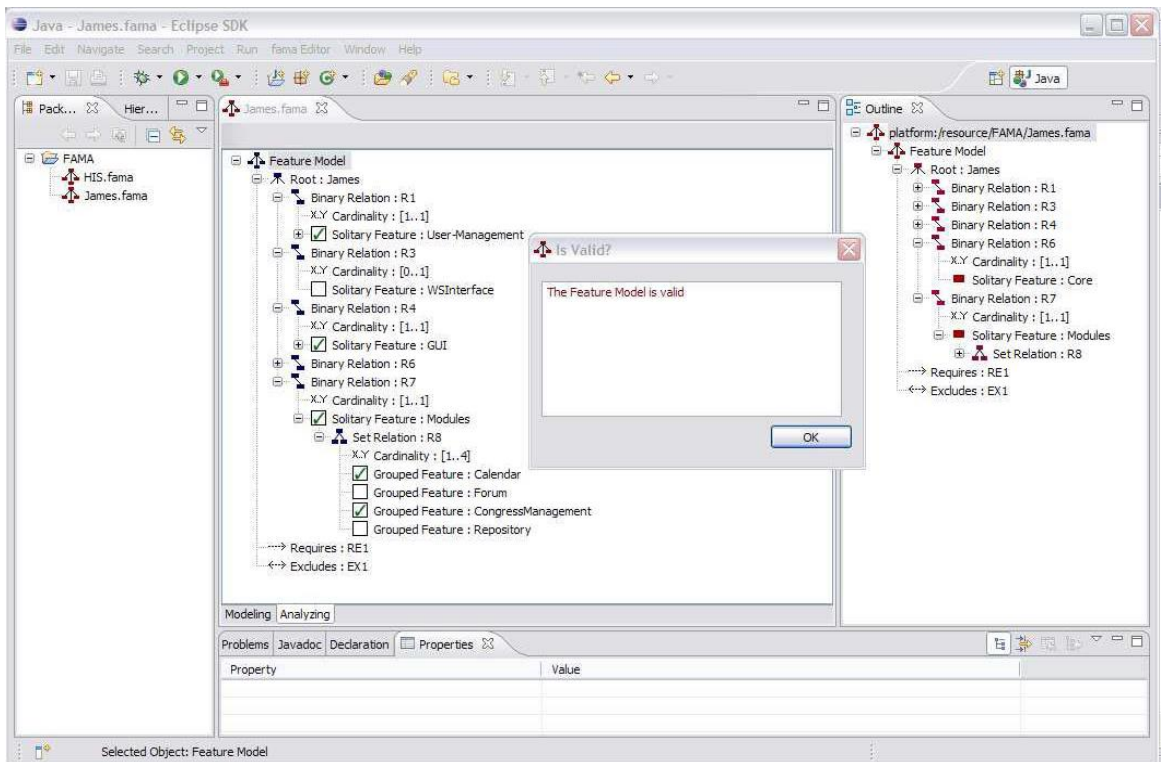

Ecplise Plugin

Here also a node represents the features, relations and cardinalities of the relations in the model. FAMA integrates different solvers in order to combine the best of all of them in terms of performance. Basically the actual framework integrates CSP solver, SAT solver & BDD, Java solver to perform the analysis tasks. When we think about FAMA one advantage is the ability to select automatically, in execution time, the most efficient solver according to the operation requested by the user. Basically FAMA (13) have two main functionalities: visual model edition/creation and automated model analysis. In this process once the user has created or imported a cardinality based feature model, the analysis capability can be used. Maximum number of operations identified on feature models are using currently implemented.

The main purpose of using FAMA to

1) Finding out if an Feature Model is valid.
2) Finding the total number of possible products of an Feature Model.
3) List all the possible products of a Feature model.
4) Calculate the commonality of a feature.

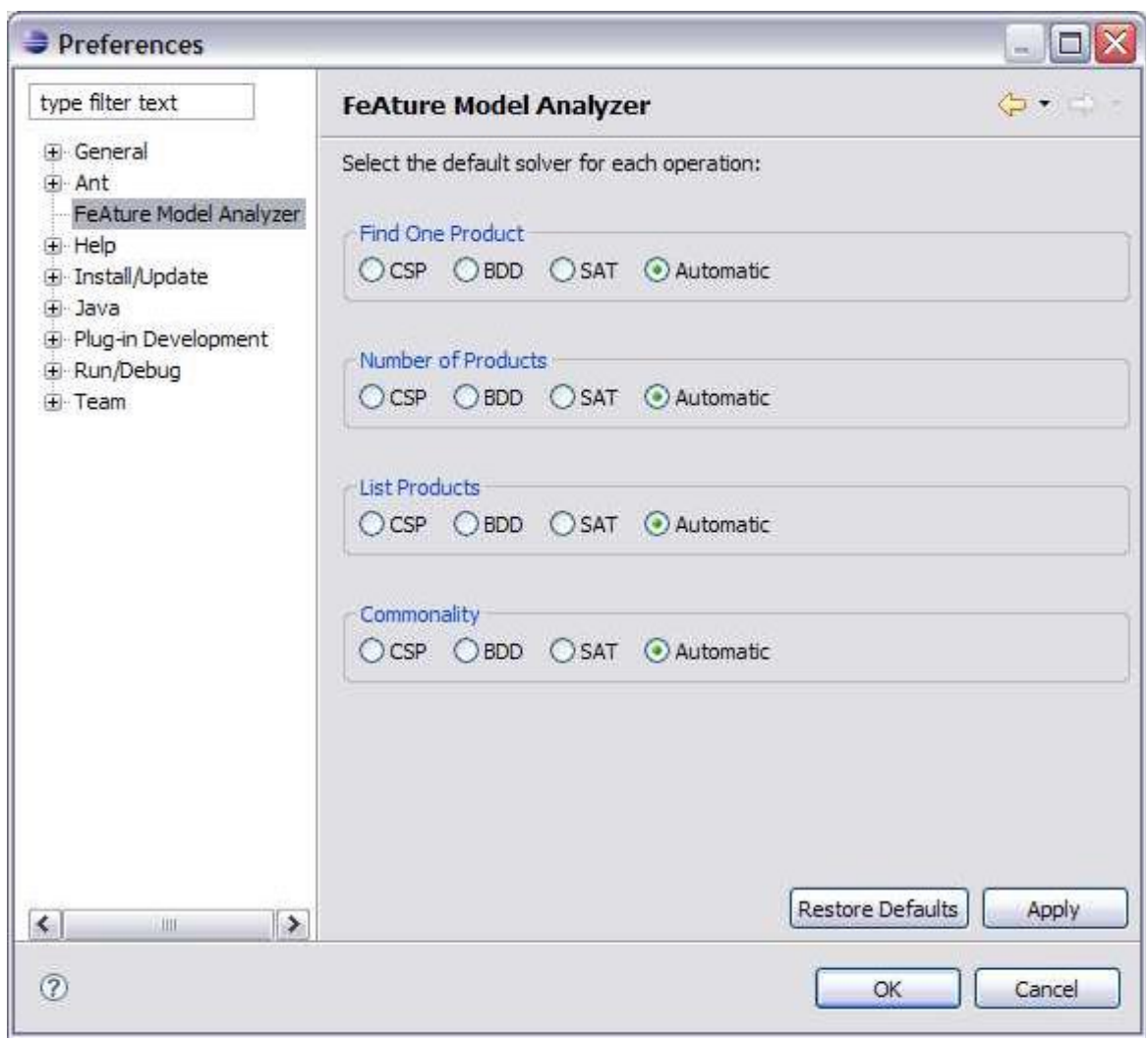

**Preference page**

In the above figure (preference page) shows a screenshot of the property page used to set the configuration options.

**11.0 Kumbang Tools:**

Kumbang Tools (17,31) is an application package consisting Kumbang configuration and Kumbang Modeller. These tools are designed for configuring the software product families. The tool takes configuration model as an input, and offers the user the possibility to make configuration decisions. The tool is implemented as plug-ins for Eclipse IDE (18).

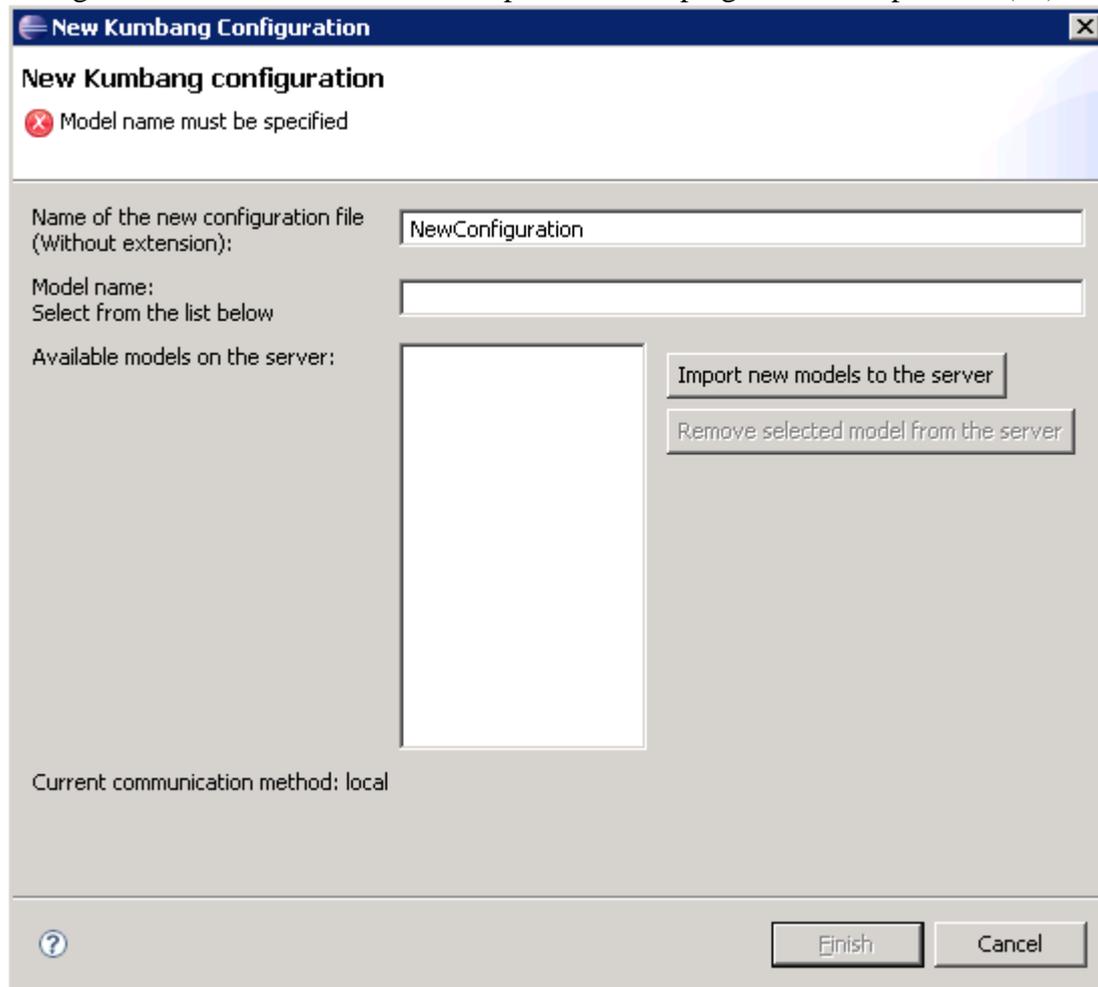

New Kumbang Configuration Dialog

Now, we are going to have a look into the Kumbang Perspective. The different views that are related to this perspective are as follows.
    a. **Features**          :      Lists the features that are defined in the model
    b. **Components**     :      Lists the components of the model.
    c. **Component diagram :**      Shows the decomposition of configuration components.

The components showed in this diagram are analogous to the components View.

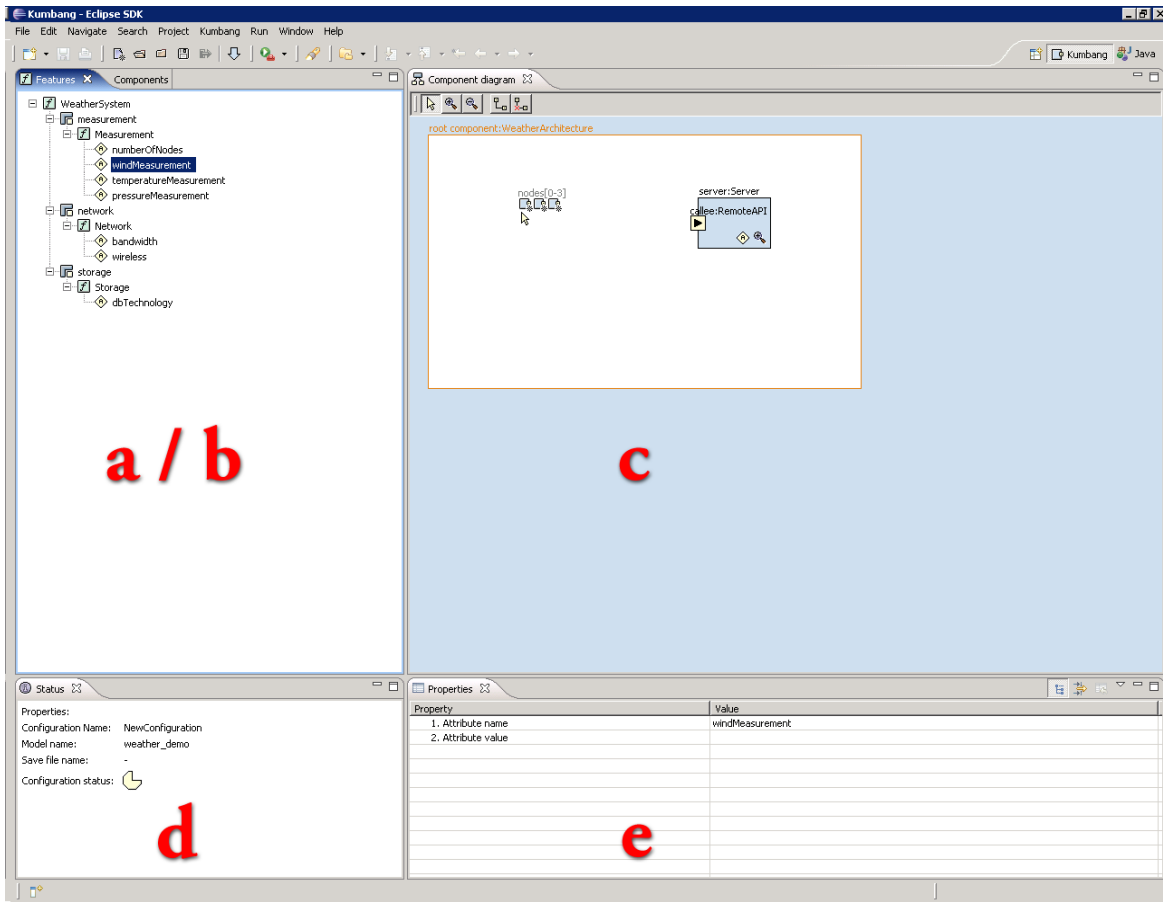

Kumbang Perspective

**d. Status**       :       shows the configuration status.

**e. Properties**   :       This show the properties are corresponding to the selected item in features/components –view

To configure the initialized model – and make it complete for exporting – you need to edit features' attributes and/or add new components to the configuration.

**Edit attributes:**
Usually the first step in modifying the configuration is editing the attributes. You can edit attributes in the Feature-view.

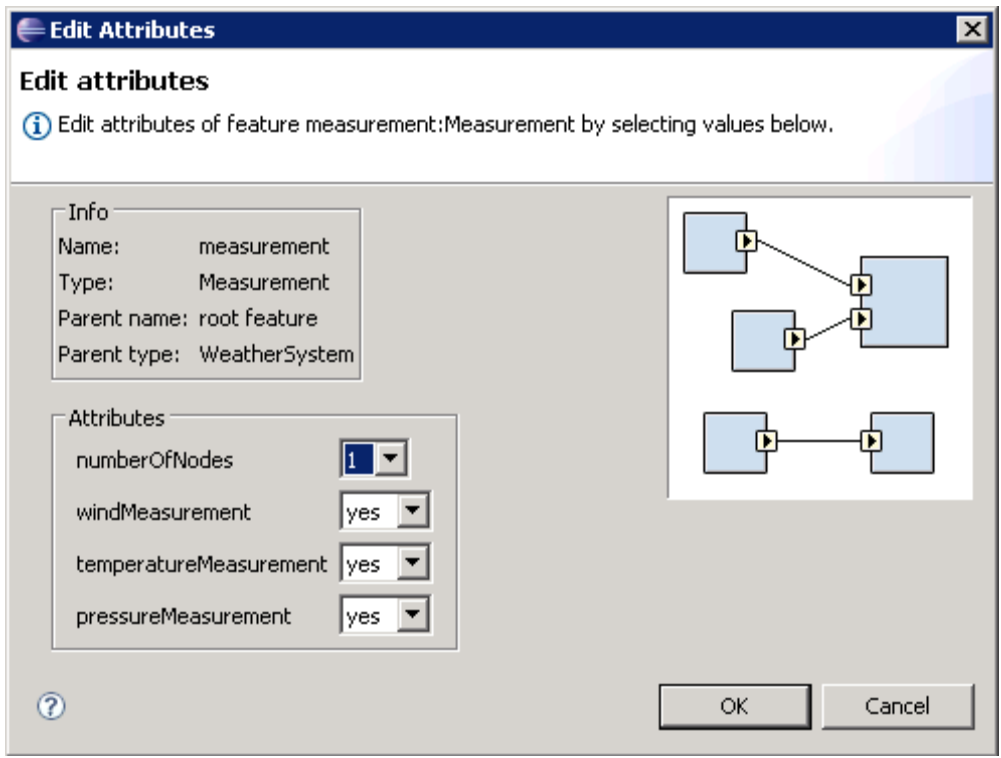

Edit Attributes Dialog (17)

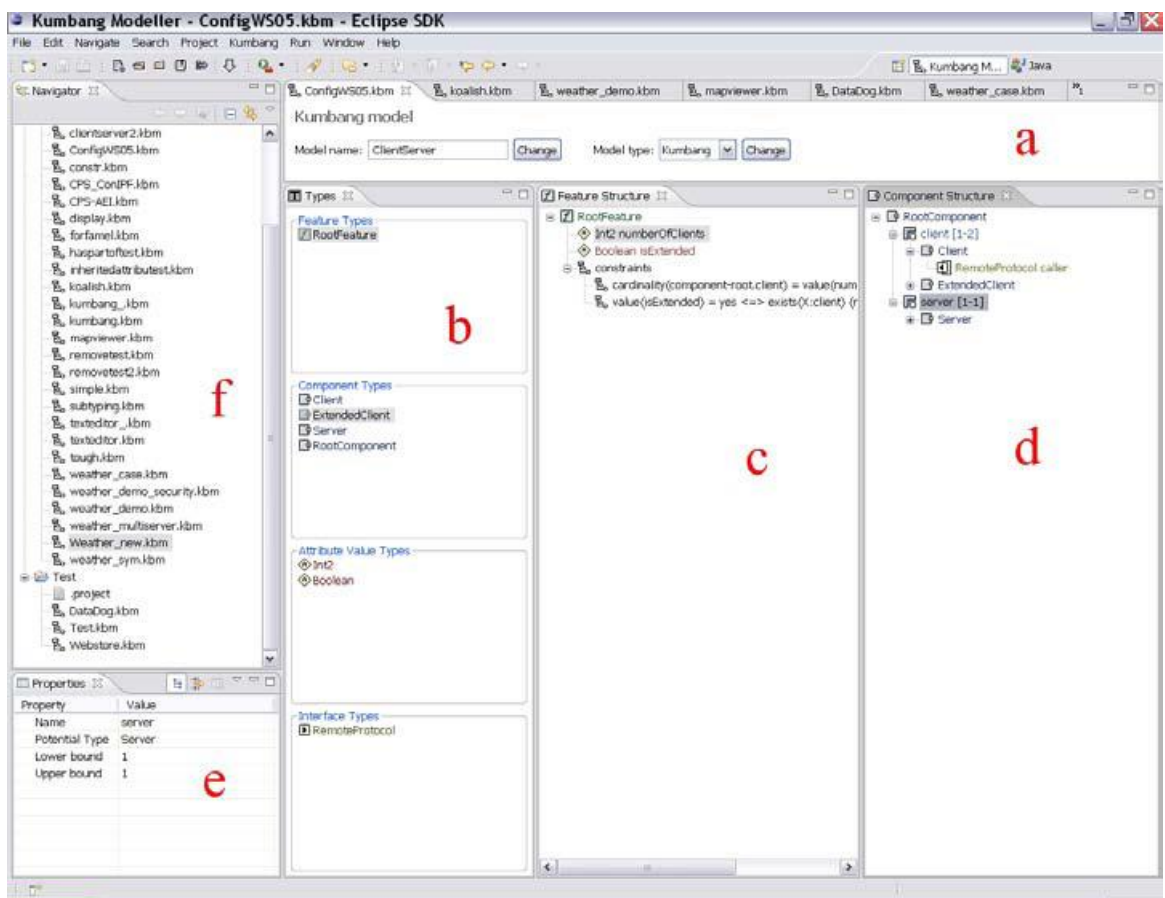

The perspective for KUMBANG modeller: a) the editor area b) the type view c) the feature view d) the component view e) the properties view f) the navigator view - fig description

## 12.0 XToF Tool:

The purpose of the XToF (10,25) tool is to let programmers define, maintain, visualise and exploit precise traceability links between a feature diagram and the code base of a software product line. Basically XTof provides enhanced functionality by leveraging on two new components 1) TagSEA, an Eclipse plug-in developed at Victoria university, which purpose is to support navigation and knowledge sharing in collaborative program development. 2) S.P.L.A.R. a Java library developed at Waterloo University that automates various FD analyses.

Below section going to present information related to the description of requirements, implementation of the initial tool chain including with its limitations. Here with this new prototype design use to overcome the aforementioned limitations. Now deeply will describe about the each one

### 12.1 Requirements:

The goal of the collaboration was to turn the implementation of a flight grade satellite communication software product line that would support the below stated requirements

- Allow *mass-customization* of the library: meaning to be able to efficiently derive products that only contain the features required for a specific space mission.
- *be compliant with quality standards and regulations* in place for flight software.
- have a *minimal impact on current development practices.*
- *Automate* the solution as much as possible.

### 12.2 The tagging languages:

Basically a feature tag is an annotation of a block of C code with the names of the features that require the block to be present. If none of the features listed in a tag is included in a particular product, then the tagged code block will not be part of the source code generated for this product. Tags can be nested and a whole file can be tagged with a special annotation. Untagged code is assumed to be needed for features.

### 12.3 Limitations of the tool-chain:

The tool-supported process described in the previous sections turned out to be effective in meeting the requirements set out by the organisation.

**12.3.1 Tighter integration:** communication between the tools was performed only through file exchange. Although this did not impede usage of the tool chain, it was recognised that an integrated environment, where loosely coupled tools play together, could be a significant enhancement.

**12.3.2 Legibility:** according to the company's developers, the legibility of the source code was not reduced by the tags. Indeed, the tagging language was designed to be concise and is rendered in a different colour in mode code editors.

**12.3.3 Portability:** although pruning dead code is most usually required in embedded systems where C dominates, C is not the only language used in embedded systems, Additionally, our "tag and prune" approach has a wider applicability than embedded systems, hence the idea of extending the approach to other languages.

**12.3.4 On-the- tag generation:** the programmers who used the tool-chain estimated that the overhead due to the tags during the domain implementation phase was 20 to 25% with respect to tag-free implementation of a 'maximal' product.

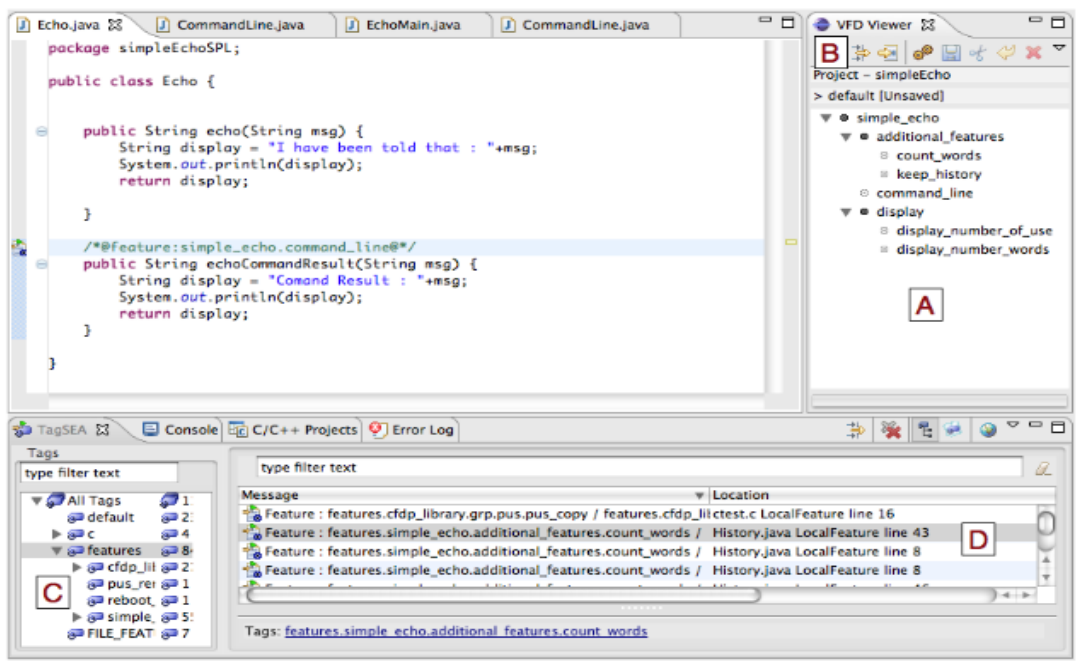

XToF's main screen

Functionally, XToF(27), the new prototype, is meant to support the activities depicted in a single integrated environment and overcome the limitations described in the previous section.

**12.4 Components and principles of XToF:**

The opportunity for re-implementing the original toolchain came from the discovery of an open source Ecplise plug-in called TagSEA. TagSEA was developed to support asynchronous and collaborative program development. It enhances navigation and knowledge distribution in the code based on tags placed by the programmers. The approach and the tool are originally unrelated to software product lines, but turned out to be applicable in this context. XToF uses the capabilities of TagSEA to manage tagging and tags. TagSEA defines waypoints as "locations of software model elements". The notation of waypoint as a point of interest has been extended to a design area of interest in order to capture blocks of code associated to feature tags. TagSEA provides mechanisms to filter tags, waypoints and navigate to a way point, XTof then links TagSEA waypoints to features and blocks of code.

## 12.5 Current functionalities:

We will have look at functionalities supported like loading the FD, tagging code fragments, navigation and visualization, configuring and pruning, improve detective tagging. Let me explain clearly each

### 12.5.1 Loading the FD:

To be displayed and configured in the tool, the FD has to be loaded. XToF expects it as an XML file in the SXFM format. The file can be created in any text editor, but can be more easily produced by the web-based visual FD editor SPLOT(26), the front-end to SPLAR. Once the FD is loaded, XToF displays it and lets the users add tags, navigate and configure. The loaded FD is copied to the project folder and its path is saved as a properly of the project. The FD is thus made available to all project contributors who can work in parallel.

### 12.5.2 Tagging load fragments:

To reduce the time needed to tag blocks of source code, XToF uses auto-completion from Ecplise. While typing a tag, feature names are displayed and when selected, directly added to the tag.

### 12.5.3 Navigation and visualization:

XToF feature tags behave like regular TagSEA waypoints. The user can list the location of feature tags. Navigate to a tagged code fragment and display it. Some visualizations have been developed to answer simple questions such as "which blocks are associated to a set of tags?" and "which set of tags is associated to a line of source code?". To answer the first question, the user can select the tags in XToF and tagged block of source code is highlighted. Another mechanism provides the opposite function i.e. answers the second question: the features corresponding to the current line in the active editor window are highlighted in the FD.

### 12.5.4 Configuring and pruning:

Configuring and pruning are now integrated. The configuration interface is based on the FD, clicking on a feature allows the user to toggle it from described to selected and conversely. Each decision made on the diagram of propagated by SPLAR to ensure the validity of the configuration is completed; the mission-specific implementation can be generated.

Code highlighting in XToF

Portability: XToF takes advantage of the plug-in platform provided by ecplise to support other languages than java. Two languages are currently supported: java and C

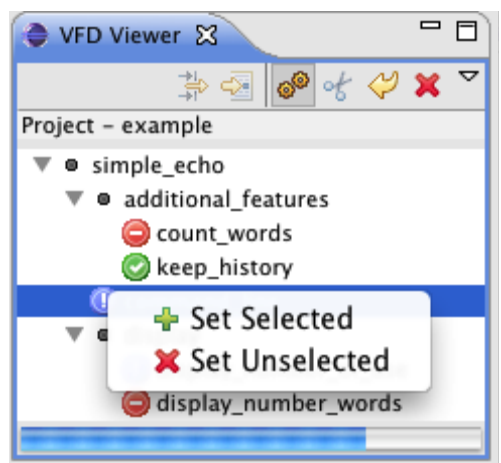

Product configuration in XToF

**13.0 PLUSEE:**
The scope of the PLUSEE (11,28) (HICSS'07) includes the product line engineering and product derivation phases

**13.1 Product line engineering:** A product line multiple-view model, which addresses the multiple views of a software product line, is modelled and checked for consistency between the multiple views. The product line multiple-view model and architecture is captured and stores in the product line reuse library.

**13.2 Product derivation:** A target system multiple view models is configured from the product line multiple-view model. The user selects the desired features for the product line member and the tool configures the target system architecture.

The PLUSEE represents second generation product line engineering tools which build on experience gained in previous research. PLUSEE builds on the experience gained with the earlier research with the knowledge based engineering environment (KBSEE). Whereas the KBSEE proof-of-concept prototype demonstrated that product line derivation from a product line feature model, architecture and components was feasible, it suffered from some serious limitations. Firstly, it used a structures analysis tool as a front end, and therefore had to rely on graphical editors for data flow diagrams and entity-relationship diagrams, which lacked the richness needed to model object-oriented product lines. Secondly, although a product line repository was used, it was developed in an ad-hoc way and lacked the underlying meta-model to formally describe the product line artefacts and their relationships. This experience with KBSEE guided the following design decisions for the development of the PLUSEE.

Both Rose and Rose RT commercial CASE Tools were used as a graphical interface to this prototype. Rose supports all the views of the standard UML notation, but it does not generate an executable architecture from the product line multiple-view model. On the other hand, Rose RT generates are executable architecture from the product line multiple-view model and simulates the product line architecture although it does not support all the views of the standard UML. To take advantages of Rose and Rose RT, two separate versions of PLUSEE, which are very similar to each other, were developed.

The knowledge based requirement Elicitation Tool (KBRET) and GUI developed in previous research were used without change. The knowledge based requirement Elicitation Tool (KBRET) is used to assist a user to select optional features of each target system. KBRET, which was developed in previous research conducts a dialog with a human target system requirements engineer, presenting the user with the optional features that will belong to the target system; KBRET.

**13.0 DecisionKing**:
DecisionKing(12) tool developed to give support for the approach integrated modelling. This application is based on the Ecplise platform.

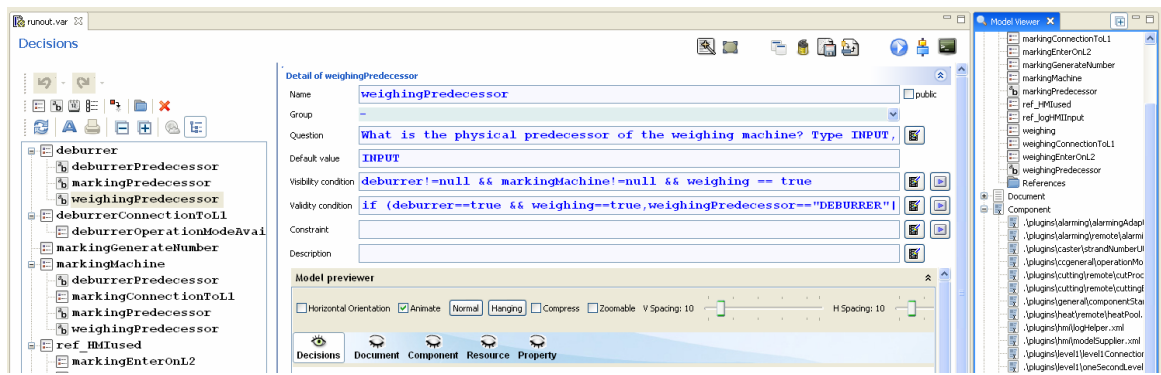

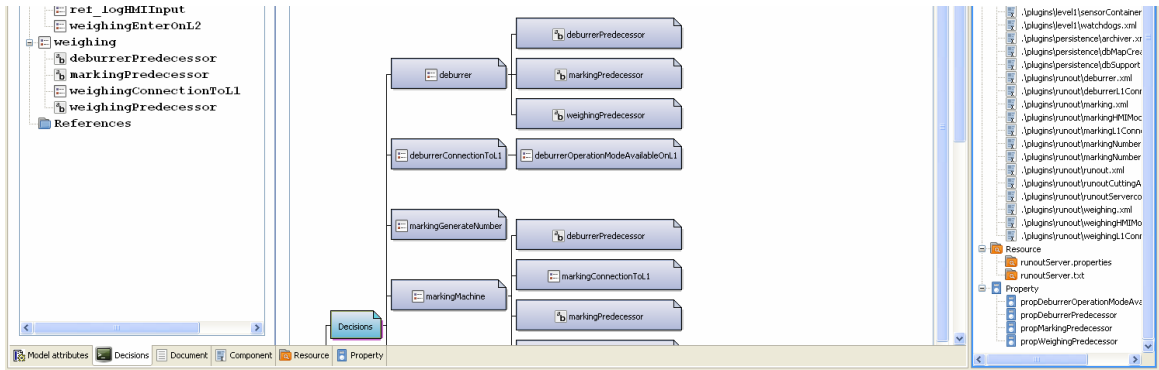

Editing a decision model in DecisionKing

The tool has been implemented in a highly iterative process with continuous feedback from Siemens VAI engineers. In the early days the versions of our modelling approach were tested using prototypes built with MS Excel. When we look into the suitability, adequacy, and usability of this approach have been tested by engineers in Siemens VAI who have been using the tool to create variability model s for different subsystems of the caster automation software. in the above figure shows a snapshot of the modelling shell in decisionKing(30). Decisions described are listed in the left pane. The right pane shows a decision viewer graphically visualizing dependencies among decisions. There are different tabs allowing importing and capturing the assets into the product line.

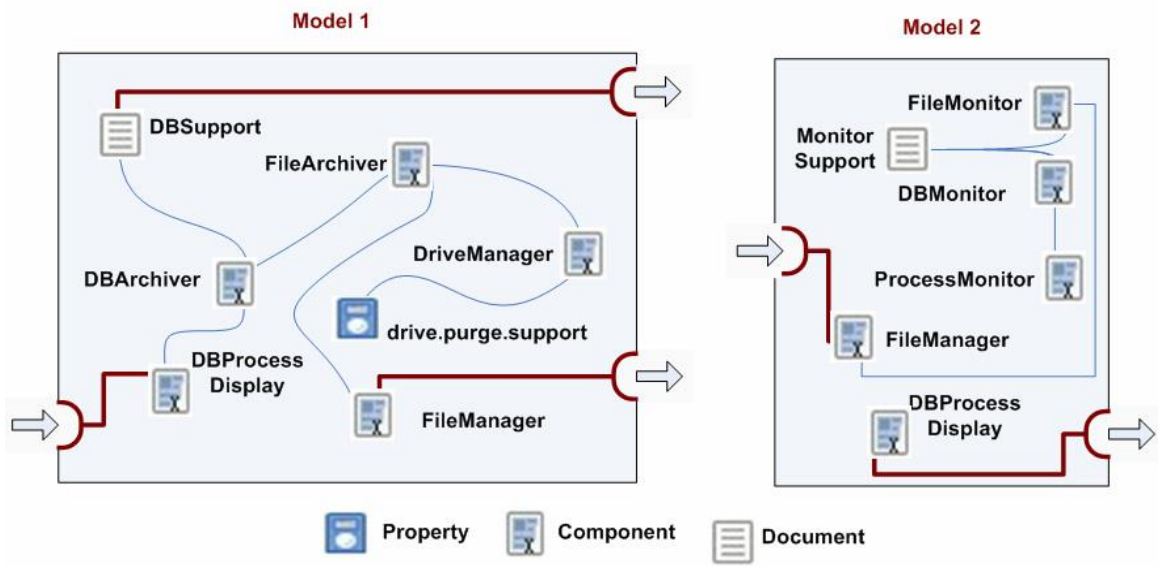

For example if we look into the above figure, the *component* tab allows to import components from existing configuration and to specify the links to the decision model. The *document* tab is used to organize fragments of the documentation. Complex relationships between decisions and assets are expressed in a simple rule language. We are replaced in the near future with an off-shelf engine to ensure scalability.

If we look into the feedback given by our industry partners' shows that the modelling approach works well for a capturing variability both from customer/marketing as well as from technical perspectives, but it is unrealistic to assume that such a model can be created and evolved by an individual or by a small team. The knowledge required to build such a model is typically spread across the minds of numerous heterogeneous stakeholders and different teams responsible for various parts of the system. DecisionKing allows modelling different parts of a large variability model separately and merging the parts into one integrated model later on. For this purpose, one team may only build the asset model, or even only a partial asset model or decision model. The different parts of the model are then merged using the model merger. Engineers can mark certain elements in the variability model as "public" meaning that these can be used in other variability models. Other elements are listed as references. A team of stakeholders responsible for a certain variability model can refer to elements of other variability models.

In the above fig we given a example depicted two parts of a variability model are shown model 1 imports DBProcessDisplay, a component defined public in model 2. Similarly model 2 refers to the FileManager component, which is set merger can combine the two models by resolving these refernces. Many problems can occur while merging different models, for example – missing refernces, multiple occurrences of the same element, or ambiguity in the mapping of referenced elements. When conflicts cannot be resolved automatically our merger relies on input from the user.

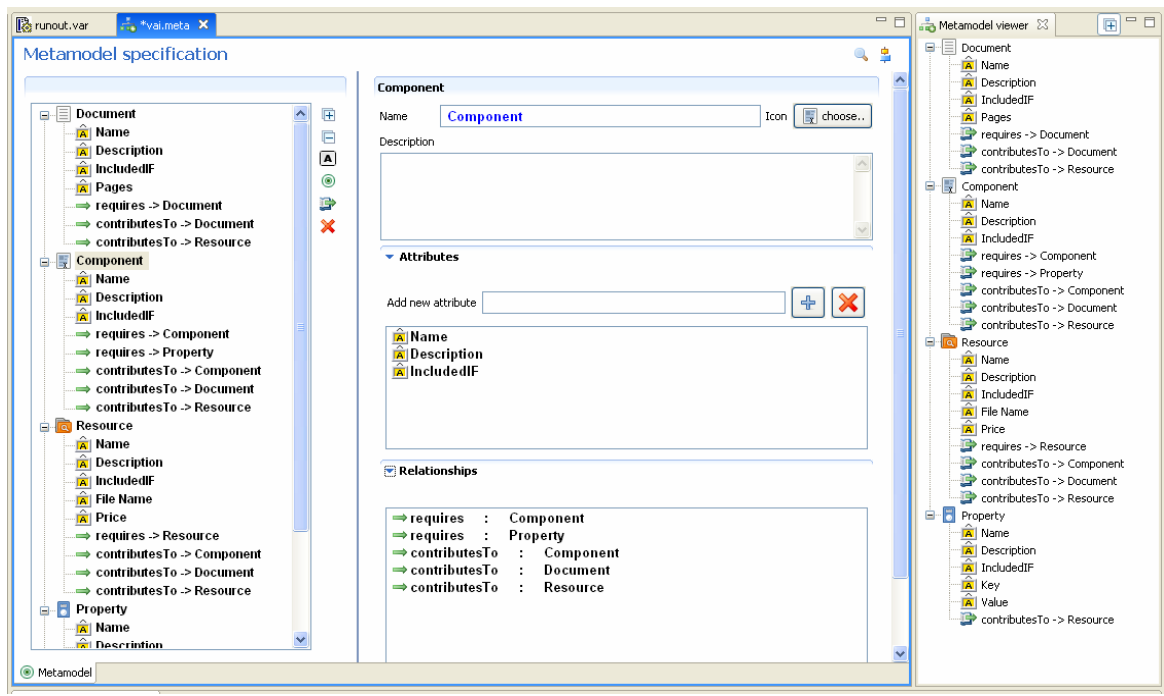
Meta-model Editor

### 14.1 Plug-in mechanism in DecisionKing:

Our second architecture-level variability mechanism ensures extensibility. DecisionKing is based on a plug-in architecture allowing arbitrary external tools to communicate and interact with it. This enables users to develop and integrate company-specific functionality. We have used this feature in three cases so far (i) we can automatically import existing assets and their relationships from existing configurations to populate the variability model. (ii) Our language to describe rules and constraints for relationships between decisions and is provided vis a plug-in. we intend to replace our current rule language with a more powerful language based on the JBoss rulw engine. (iii) we can use third-party model differences as demonstrated with a plug-in(29) by we integrated via this mechanism.

**15.0 BVR TOOL (Base-variation-resolution):**

The BVR (14) (NIK'06) approach depends on the possibility of establishing and maintaining the relations between the variation models and the base model, and between the resolution and variation models. In order to explain this concept, we have built a prototype tool called Object-Oriented Feature Modeller (OOFM). For the purpose of this prototype tool implementation we have used Java as the language of the base models – it could just as well have been like UML.

Initially we will know about BVR approach, this approach is defined by a meta model divided into three parts. The base model will be any model in a given language. The variation model will contain variation Elements, where each element refers to the base model Element this is subject to variation (implying that those that are not related are not subjected to relation). This relationship has a zero-to-one cardinality, as not all model Elements are affected by variability, variation elements only contain the information that the referenced model elements may be affected by variations; the information contained in the base model element is not duplicated. Variation is specified in a variability specification; it may in general involve other model elements and affect a number of variation elements. Variability specification comes in two kinds: variability constraint represents constraints on valid resolutions and distinguishes between valid resolution models and invalid ones; Transformers have concrete transformation associated with them. When values are bound to transformers (from the resolution element), this defines the transformation of the variation model and the base model into a specific model.

The OOFM prototype tool was made in parallel with the development of the BVR approach. Therefore its variation model has a slightly different set of meta classes. According to the variation model of the OOFM, a model contains exactly one product, product may have zero or many features, each containing zero or many features. That is, a feature can contain other features. Waterproof is an example of a feature that contains two other features: depth and Time. Feature cardinality is represented as mandatory ([1….1]), optional ([0….1)] and group feature cardinality as alternative (<1-n>). A feature is mandatory, optional or alternative. Feature choices are stored in a List in the container feature object. For example the sub-feature Depth has the choices 50 and 100, which are kept in the choice List in the feature object Depth. Depth choices state the waterproof depth (in meters) of a watch. Similarly, the sub-feature Time has a choice List that contains the choices 0, 5, 10 and 15. Time choices (in hour) tell us how many hours a waterproof watch can be under water before it no longer can resist water. The links between Variation model and Base model indicate to which element of the Base model the variation applies. Feature definition in OOFM is not totally automatic. OOFM has the ability to recognize and display all object fields it can define as features. The resulting resolution model will contain the variable features from the variation model and those objects fields of the base model that were not defined as features.

To implement this tool, this is made as an Ecplise plugin and based upon the ECplise modelling framework (EMF). The feature modelling editor is based upon a Meta model according to the variation model part of the Meta model. This is done by defining the Meta model in terms of annotated Java classes and using the generator for tree-oriented model editors provided by the EMF. The Java development technology (JDT) is used to represent

Base Java programs and Java programs that are generated Java programs in terms of objects according to JDT.

# 16.0 Comparative Results & Analysis of Variability Management Tools

## 16.1 VISUALIZATION TECHNIQUES:

| VISUALISATION TECHNIQUES |||
|---|---|---|
| **Tools** | **Approach & Area** | **Visualisation Type** |
| Pure:Variants | Product line engineering | Graph Representation, Table, Box & Arrow, Textual |
| KUMBUNG | KUMBANG | GUI TREE, Box & Arrow, Textual Language |
| AHEAD | FOP (Feature Oriented Programming) | GUI TREE, TABLE, TEXTUAL |
| FAMA | FAMA | GUI TREE, TEXTUAL LANGUAGE |
| XToF | Tagging | GUI TREE, TEXTUAL |
| CONSUL | CONSUL | GUI TREE |
| COVAMOF | COVAMOF | CVV (COVAMOF Variability view) |
| PLUSEE | PLUSEE | GUI TREE |
| BVR Tool | BVR approach (Base-Variation-Resoluton) | GUI TREE, TABLE, BOX & Arrow, Textual |
| Feature Modelling Tool | FODA | GUI TREE, TEXTUAL LANGUAGE |
| GEARS | Software Product Line | GUI TREE, TEXTUAL LANGUAGE |
| Feture Plug-in for eclipse | FODA | GUI TREE, TEXTUAL LANGUAGE |
| DecisionKing | Integrated Modelling | GUI TREE |
| VMWT | Web Based | GUI TREE, TABLE, TEXTUAL LANGUAGE |

From the above table information about tools visualization techniques and from which approach that is tool is retrieved. All the tools have some type of visualization technique like tree, table, textual and box & arrow. Each and every tool must use any of these or their own technique to visualize features at the same way addition to this information in the above table approaches also mentioned if available from the tool retrieved.

## 16.2. FUNCTIONAL CRITERIA:

| FUNCTIONAL CRITERIA | | | | | |
|---|---|---|---|---|---|
| **Tools** | VBT | FIT | NF | AFN | FC |
| Pure:Variants | PS | PS | PS | PS | PS |
| KUMBUNG | PS | NS | NS | NS | FS |
| AHEAD | PS | NS | NS | NS | FS |
| FAMA | FS | NS | NS | NS | PS |
| XToF | PS | NS | PS | NS | NS |
| CONSUL | PS | PS | PS | PS | PS |
| COVAMOF | FS | NS | NS | NS | FS |
| PLUSEE | PS | PS | PS | NS | NS |
| BVR Tool | PS | PS | NS | NS | PS |
| Feature Modelling Tool | PS | NS | NS | NS | PS |
| GEARS | FS | NS | NS | PS | FS |
| Feture Plug-in for eclipse | PS | NS | NS | NS | PS |
| DecisionKing | PS | PS | NS | NS | FS |
| VMWT | PS | NS | NS | NS | NS |
| | | | | | |
| VBT: Variability Binding Time | | FS: Fully Supported | | | |
| FIT: Feature Implementation Time | | PS: Partially Supported | | | |
| NF: Negative Features | | NS: Not Supported | | | |
| AFN: Alternative Feature Names | | | | | |
| FC: Feature Cardinality | | | | | |

**FUNCTIONAL CRITERIA:** In this section of this work the functional criteria of each Tool is covered. This criterion addresses the characteristics and attributes of features and variation points that should be addressed and captured within the variability model.

**16.2.1 VBT (Variability Binding Time):** In previous works variation points are places in the design or implementation where variation occurs. Variability is due to unmade decisions that are left open as long as economically feasible. Anyway, specifying the point in time when a variation point is to be bound to a specific variant is important. A number of possible binding times have been identified and used in industry.

**16.2.2 FIT (Feature Implementation Time):** Current industry software systems are usually built incrementally, there is a rarely a software product that is built as a final release from the first edition. Products are usually enhanced and features added to them continuously over time. Planning for further releases of products, the features to be implemented in these products, and the timing, is a key step for the success and sustainability of a product line. The feature implementation time should be captured with in the variability model as it contributes to the process of product versioning.

**16.2.3 N.F (Negative Features):** Basically, the development of variability models has purely depended on the features that are to be supported by a product line. At the same time little attention paid on the features which are not supported. These product ranges from low-end products to high-end ones. Negative features are features that are not supported by the given products. In such cases the product architecture should be designed in way to prohibit the enabling of such features by end user of the product.

**16.2.4 AFM (Alternative Feature Names):** In a software product line life cycle there are so many areas variability management exits from requirements, to architecture design and implementation. Different people will use different ways to find out the variability and to express features. So, in this case the same feature may have different names in another team that need to watch carefully.

**16.2.5 FC (Feature Cardinality):** Upto economically feasible always desirable to delay design decisions. One potential solution to alleviate the effect of open variation points is by attaching a limited number of possible variants that could be bound to a given variation point. This is usually referred to as feature cardinality.

## 16.3 NON-FUNCTIONAL CRITERIA:

| NON-FUNCTIONAL CRITERIA | | | | |
|---|---|---|---|---|
| Tools | Functional Dependency | Functional Interaction | Tangled features | Behavioural features |
| Pure:Variants | FS | PS | PS | NS |
| KUMBUNG | FS | PS | PS | PS |
| AHEAD | FS | FS | NS | NS |
| FAMA | FS | PS | PS | NS |
| XToF | FS | PS | NS | NS |
| CONSUL | FS | FS | PS | NS |
| COVAMOF | FS | NS | NS | NS |
| PLUSEE | FS | PS | PS | PS |
| BVR Tool | FS | PS | NS | NS |
| Feature Modelling Tool | FS | PS | NS | NS |
| GEARS | FS | PS | FS | NS |
| Feture Plug-in for eclipse | NS | PS | NS | NS |
| DecisionKing | FS | FS | NS | NS |
| VMWT | FS | PS | NS | NS |
| | | | | |
| | | FS: Fully Supported | | |
| | | PS: Partially Supported | | |
| | | NS: Not Supported | | |

**16.3.1 NON-FUNTIONAL CRITERIA:**
This section represents the relationship between two or more features. These relationships are classified based on their type and how they affect other features within the variability model as well as the system architecture.

**16.3.2 F.D (Feature Dependencies):** In the same feature model, features in a model affect each other in a number of ways. Some features cannot be supported unless other features are supported in a product; other features may conflict and cannot be supported in the same product at the same time. Other forms of dependency could include weaker from of relationships such as when the inclusion of some feature recommends the inclusion/exclusion of another. Dependencies can be quite difficult to model especially those that relate to quality attributes. Hence, dependencies should not only be represent as first class citizens in any variability model, but also the technique used for capturing dependencies should allow for complex dependency representation.

**16.3.3 F.I (Feature Intereaction):** In feature models with some presence and absence of features it may afect the other features, feature interaction is concerned with how different feature combinations affect the system architecture. Different feature combinations might lead to the inclusion of different architectural components and configurations.

**16.3.4 T.F (Tangled Features):** The phase in software product line is the mapping of the selectable and configurable features to their corresponding implementation components. This process of encapsulation of features exhibiting non-functional properties is often limited due to their crosscutting nature. The way for deal with cross cutting features is Aspect-oriented Development(AOD). This will allow isolating and thereby encapsulating the implementations of crosscutting concerns in class like modularization units called aspect.

**16.3.5 B.H (Behavioral Features):** This is one of the crucial part of the management level of the variability model. It is well known as capturing behaviour. This is due to the fact that some variability requirements encompass behavioural information that could not to be captures using traditional approaches. Another example is capturing information relating to data flows and data paths. Many approaches have been proposed to capture behaviour, from using UML state charts and use case diagrams within the multiple-views of the variability.

**16.4 GOVERENCE ISSUES:**

| GOVERNANCE ISSUES | | | | |
|---|---|---|---|---|
| Tools | Cost Benefit Analysis | Open & Closed Set of Features | Multiple Views | Multiple Users & Access Control |
| Pure:Variants | NS | NS | PS | PS |
| KUMBUNG | NS | NS | PS | NS |
| AHEAD | NS | NS | NS | NS |
| FAMA | NS | PS | PS | NS |
| XToF | NS | NS | PS | NS |
| CONSUL | NS | NS | NS | NS |
| COVAMOF | NS | PS | PS | NS |
| PLUSEE | NS | NS | PS | NS |
| BVR Tool | NS | NS | PS | NS |
| Feature Modelling Tool | NS | NS | NS | NS |
| GEARS | NS | PS | NS | NS |
| Feture Plug-in for eclipse | NS | NS | NS | NS |
| DecisionKing | NS | NS | PS | NS |
| VMWT | FS | NS | NS | NS |

| | |
|---|---|
| FS: Fully Supported | |
| PS: Partially Supported | |
| NS: Not Supported | |

**16.4.1 GOVERENCE ISSUES:**
This section will deal with business concerns of the software product line in general as well as the construction and management of the variability model.

**16.4.2 C/B Analysis (Cost/Benefit):** To find and document the cost available in the overall project including valuable input. The cost for realizing a feature could be captured in the form of a financial estimate or man/month effort needed. The benefit could range from allowing for lower implementation costa and faster time-to-market to enhancing market shares and increasing the competitive edge of product line. Generally it is not an easy task to specify the cost/effort and benefit involved in realizing a given feature, adequate estimates can be obtained using information gathered and experiences gained from previous similar projects.

**16.4.3 O/C.S.F (Open/Closed set of features):** Inside the industry projects very hard for the architect to built with a comprehensive and complete set of features. Rather than that features are continuously added to the initial feature model over time – even after the design process is completed. It is very hard to design a system that has around an open and changing set of features that can be modified anytime. To overcome this problem, some industries differentiate between two types of features. The features which are cannot be changed or modified by the architect or development team and serve as the core of the product or product line. The features can be able to change or alter with advance in technology that to with out effecting to the overall system are open features. Such features can be altered by project manager, architect, or the development team depending on the nature of the feature.

**16.4.4 M.V (Multiple Views):** Mostly different stakeholders have interest in considering different views of the product line variability model. So it is very important to present the extract information in multiple views for different groups of stakeholders like users, system analysts, developers, etc. The main challenge of multiple views is preserving consistency. So, for this purpose introduce meta-views to check for inconsistencies.

**16.4.5 M.U & A.C (Multiple Users & Access Control):** As per the above multiple views, each view will be targeted at a specific user group. It is very important that a variability management solution provides access control to the variability model data. So in this way the user can only see information relevant to their view and can only modify properties that are within their limit.

**16.5 Technical Aspects**

| TECHNICAL ASPECTS ||||| 
|---|---|---|---|---|
| Tools | Traceability | Estimation of Number Of Products | Phases of Life Cycle | Tool Approach |
| Pure: Variants | NO | NO | Design, Analysis, Implementation | Configuring & Modelling |
| KUMBUNG | NO | YES | Implementation | Configuring |
| AHEAD | YES | YES | Design, Implementation | Configuring & Modelling |
| FAMA | NO | YES | Implementation | Configuring & Modelling |
| XToF | YES | NO | Implementation | Configuring |
| CONSUL | NO | NO | Implementation | Configuring & Modelling |
| COVAMOF | YES | NO | Design, Analysis, Implementation | Management |
| PLUSEE | NO | NO | Analysis | Configuring & Management |
| BVR Tool | NO | NO | Implementation | Configuring |
| Feature Modelling Tool | NO | NO | Design | Configuring |
| GEARS | YES | YES | Analysis, Design | Modelling, Configuring & Management |
| Feature Plug-in for eclipse | NO | NO | Design | Configuring & Modelling |
| DecisionKing | YES | NO | Implementation | Modelling |
| VMWT | NO | YES | Design, Implementation | Configuring, Modelling & management |

## 17. Conclusion

This survey contains the tools which deals with variability management in software product lines. This report included different tools from various approaches and their working conditions. Functional criteria contains the information of the tools regarding to the functional strategies whether supporting or not supporting, at the same way non-functional criteria. Visualization area deals with visualization type of the tools and type of representation e.g. tree type, intended list, graphical. Goverence issues deals the supporting nature of the tools towards goverence issues stated in the report. Technical area gives us the information of the tools working nature and tooling approach. With this information about the tools industrial people can pick their tools easily. Tools including in this report are covering tooling approaches like modeling, configuring and management. Information about the covering phases of software product line life cycle of tools also presented.

## References

1) Rafeal Capilla, Sanchez.A, Juan C . Duenas: An analysis of variability modeling and management Tools for product line development. : In proceedings from workshop on variability management, Finland, PP-36-38, 2007.
2) Rafeal Capilla, Sanchez.A, Juan C . Duenas: An analysis of variability modeling and management Tools for product line development. : In proceedings from workshop on variability management, Finland, PP-39-40, 2007.
3) Rafeal Capilla, Sanchez.A, Juan C . Duenas: An analysis of variability modeling and management Tools for product line development. : In proceedings from workshop on variability management, Finland, PP-40, 2007.
4) Rafeal Capilla, Sanchez.A, Juan C . Duenas: An analysis of variability modeling and management Tools for product line development. : In proceedings from workshop on variability management, Finland, PP-41, 2007.
5) M. Fritzsche, W. Gilani, I. Spence, P. Kilpatrick, TJ Brown, and R. Bashroush. "Towards Performance Related Decision Support for Model Driven Engineering of Enterprise SOA Applications." Proceedings of the 15th IEEE International Conference on Engineering of Computer-Based Systems (ECBS), Belfast, Northern Ireland, April 2008.
6) C. Gillan, P. Kilpatrick, I. Spence, R. Gawley, T.J. Brown and R. Bashroush. Challenges in the Application of Feature Modelling in Fixed Line Telecommunications. Proceedings of the First International Workshop on Variability Modelling of Software-intensive Systems (VaMoS2007), Lemrick, Ireland, Jan 16 -18, 2007.
7) R. Bashroush, I. Spence, P. Kilpatrick, and TJ Brown. Deriving Product Architectures from an ADLARS Described Reference Architecture using Leopard. ACM/SIGSOFT Foundations of Software Engineering FSE-12, Newport Beach, California, October 2004.
8) R. Bashroush. "A NUI Based Multiple Perspective Variability Modelling CASE Tool," Muhammad Ali Babar, Ian Gorton (Eds.): ECSA 2010. Lecture Notes in Computer Science, Volume (6285), Springer-Verlag Berlin Heidelberg, ISBN 978-3-642-15113-2, August 2010.

9) R. Bashroush. "A Scalable Multiple Perspective Variability Management CASE Tool". Proceedings of the 14th International Software Product Line Conference (SPLC), South Korea. September 2010.
10) R. Bashroush, R. Perrott. Using a Software Product Line Approach in Designing Grid Services. Proceedings of the 4th UK e-Science AHM2005, Nottingham, UK, September 2005.
11) R. Bashroush, I. Spence, P. Kilpatrick, and TJ Brown. "Towards More Flexible Architecture Description Languages for Industrial Applications," V. Gruhn and F. Oquendo (Eds.): EWSA 2006, Lecture Notes in Computer Science, Volume (4344), pp. 212-219. Springer-Verlag Berlin Heidelberg, 2006.
12) Danilo.Beuche, Holger.Papajewski, December 2004, Variability management with feature models, Science of computer programming, vol.53,no.3, pp.333-352.
13) T.J. Brown, R. Bashroush, C. Gillan, I. Spence, and P. Kilpatrick. Feature Guided Architecture Development for Embedded System Families. Proceedings of the 5th Working IEEE Conference on Software Architecture WICSA. Pittsburgh, PA, USA, November 2005.
14) T.J. Brown, R. Gawley, I. Spence, P. Kilpatrick, C. Gillan and R. Bashroush. Requirements Modelling and Design Notations for Software Product Lines. Proceedings of the First International Workshop on Variability Modelling of Software-intensive Systems (VaMoS2007), Lemrick, Ireland, Jan 16 -18, 2007.
15) Jose Evelio Martinez Saiz, Feature Models Visualization Based on Ontology Framework, Thesis submitted to university of Vrije, Brussel, PP- 22-23, 2008-09.
16) Jose Evelio Martinez Saiz, Feature Models Visualization Based on Ontology Framework, Thesis submitted to university of Vrije, Brussel, PP- 23-25, 2008-09.
17) Jose Evelio Martinez Saiz, Feature Models Visualization Based on Ontology Framework, Thesis submitted to university of Vrije, Brussel, PP- 25-26, 2008-09.
18) Jose Evelio Martinez Saiz, Feature Models Visualization Based on Ontology Framework, Thesis submitted to university of Vrije, Brussel, PP- 27-28, 2008-09.
19) Chirstophe. Gauthier, XToF- A Tool for tag-based Product line Implementation, PReCISE Reasearch Centre, Namur, Belgium, 2009.
20) Gomaa.H, Michael E. Shin, Automated software Product line Engineering and Product Derivation,: Proceedings of 40[th] Annual Hawalii International Conference on system Sciences (HICSS'07), 2007. IEEE
21) Deepak.D, Paul.G, Rick Rabiser, DecisionKing : A Flexible and Extensible Tool for Integrated Variability Modeling, Johannes Kepler University, 4040Linz, Austria.
22) D.Benavides, S.Segura, P.Trinidad, Antonio R.Cortes. FAMA: Tooling a Framework for the Automated Analysis of Feature Models. In Proceecings of First International conference on Variability Modelling, 2007.
23) P.Shakari, Birger M.Pederson. On the implementation of a Tool for Feature modelling with a Base Model Twist. Proceedings of 'NIK' conference, 2006.
24) "BigLever Software GEARS", http://www.biglever.com/extras/splLifecycleFramework.PDF.
25) M.Antkiewicz, K.Czarnecki. Feature Plugin: Feature Modeling plug-in for Ecplise. In: Proceedings of the 'OOPSLA' workshop on Ecplise Technology ExChange(ETX), oct 24-28, Vancouver British Columbia, Canada, 2004.
26) V. Myllarniemi, M.Raatikainen, P. Lahti. Kumbang Tool's User's Guide, Proceedings of Software product Line Conference, august, 2007.
27) K. Pohl, P. Heymans, KC. Sang. Kumbang Sec: An approach for Modelling Functional and security variability in software Architecture. Proceedings of VaMos'07, Limerick, Ireland, Janvary 16-18, 2007.
28) 'Pure-systems GmbH', *'pure::variants community variant management',* retrived on 10[th] December 2010 from http://www.pure-systems.com

29) 'AHEAD', 2008, *The AHEAD Tool Suite(ATS)',* Retrived on 11[th] December 2010 from http://www.cs.utexas.edu
30) Grupo de Investigacion en Reutilizacion y Orientacion a Objeto (GIRO) – *Feature Modeling Tool.* Avaliable from: http://www.giro.infor.uva.es/FeatureTool.html
31) Ruben Fernandez, Miguel A. Laguna, Jesus Requejo, Nuria Serrano (2009). *Development of a Feature Modeling Tool using Microsoft DSL Tools.* Department of Computer Science, University of Valladolid.
32) Pure-systems GmbH (2008). *Pure::varints User's Guide: Version 3.0 for pure::varints 3.0.* Avaliable from: http://www.pure-systems.com/fileadmin/downloads/pure-varints/doc/pv-user-manual.pdf
33) Feature Modeling Plug-in. Generative Software Development Lab – University of Waterloo: http://gsd.uwaterloo.ca/Projects/fmp-plug-in/.
34) P. Ebraert, A. Classen, P. Heymans, and T. D'Hondt, "Feature diagrams for change-oriented programming," in proceedings of ICFI'09, 2009.
35) M. Mendonca, M. Branco, and D. Cowan, "S.P.L.O.T,: Software product lines online tools," in proceedings of OOPSLA'09, 2009.
36) Q. Boucher, A. Classen, P. Heymans, A. Bourdoux, and L. Demonceau, "Tag and prune: A pragmatic approach to software product line implementation," PReCISE Research Centre, Univ. of Namur, Tech. Rep., 2009.
37) Reda Bendraou, Marie-pierre Gervals, and Xavier Blanc, "UML4SPM: A UML2.0-Based Metamodel for Software process Modeling," ACM/IEEE 8[th] International Conference on Model Driven Engineering Languages and Systems, UMl 2005, Montego Bay, Jamaica, October 2-7, 2005.
38) M. Abi-Antoun, J. Aldrich, N. Nahas, B. Schmerl, and D. Garlan, "Differencing and Merging of Architectual Views," Proceedings of the 21[st] *IEEE International conference on Automated Software Engineering(ASE'06),* Tokyo, Japan, 2006, pp.47-58.
39) K. Schmid and I. John, "A Customizable Approach to Full-Life Cycle Variability Management," *Journal of the science of computer programming, Special Issue on Variability management,* Vol. 53(3), pp. 259-284, 2004.
40) T. Asikainen, T. Mannisto, and T. Soininen, Kumbang: A domain ontology for modeling variability in software product families, *Advanced Engineering Informatics,* Vol. 21, pp. 23-40, 2007.
41) M. Sinnema and S. Deelstra, Industrial validation of COVAMOF, *Journal of systems and software,* vol. 81, pp. 584-600, 2008.
42) D. Beuche, H. Papajewski, and W. Schroder-Preikschat, Variability management with feature models, *Sci. Comput.Program.,* vol. 53, pp. 333-352, 2004.
43) Batory, D., Sarvela, J.N. and Rauschmayer, A. Scaling Step-Wise Refinement. IEEE TSE 30(6) 355-371, (2004)
44) Diaz, O., Trujillo, S. and Anfurrutia, F.I. Supporting production Strategies as Refinements of the Production Process. Proceedings of 9[th] Software Product Line Conference (SPLC), Springer-Verlag LNCS 3714 pp. 210-221, (2005).
45) Trujillo, S., Batory, D. and Diaz O. Feature Refactoring a Multi-Representation Application into a product Line. Proceedings of 5[th] International Conference on Generative Programming and Component Engineering, pp. 191-200 (2006).